# CORRELATION BETWEEN TRANSPORT PROPERTIES AND LATTICE EFFECTS IN THE $NdCoO_3$ BASED CATALYSTS AND SENSOR MATERIALS

Cristina Tealdi[1,*], Lorenzo Malavasi[1,2], Fabia Gozzo[3], Clemens Ritter[4], Maria Cristina Mozzati[5], Gaetano Chiodelli[2], Giorgio Flor[1]

[1]Dipartimento di Chimica Fisica "M.Rolla", Università di Pavia, Viale Taramelli 16, 27100 Pavia, Italy
[2]CNR - IENI Sede di Pavia, Viale Taramelli 16, 27100 Pavia, Italy
[3]Paul Scherrer Institute, Swiss Light Source, 5232 Villigen PSI, Switzerland
[4] Institute Laue-Langevin, Boite Postale 156, F-38042 Grenoble, France.
[5]CNISM, Unità di Pavia and Dipartimento di Fisica "A. Volta", Università di Pavia, Via Bassi 6, I-27100, Pavia, Italy.

*Corresponding Author: Dr. Cristina Tealdi, Dipartimento di Chimica Fisica "M. Rolla", Università di Pavia, V.le Taramelli 16, I-27100, Pavia, Italy. Tel: +39-(0)382-987921 - Fax: +39-(0)382-987575 - E-mail: cristina.tealdi@unipv.it

# ABSTRACT

This study presents correlations between the structural and transport properties of pure and doped neodymium cobaltate, a compound of great interest for its foreseen applications as catalyst, sensor and thermoelectric material.

Neutron and x-ray powder diffraction data have been combined to carefully determine lattice constants and atomic positions and four probe direct current conductivity and thermoelectric power measurements allowed us to follow the thermal evolution of the transport properties of these compounds.

The dramatic improvement of the room temperature conductivity of $Nd_{0.8}Ca_{0.2}CoO_3$ with respect to the pure and the Na-doped compound is explained in terms of a different spin-state for the Co ions within this structure. The higher conductivity and the absence of anomalies in the thermal expansion makes the Ca-doped compound more attractive than the pure $NdCoO_3$ in view of possible applications. The experimental data and the Co environment analysis here discussed, in particular bond lengths distortion and bending angles, are fully consistent with a spin state (low to intermediate) transition in $NdCoO_3$.

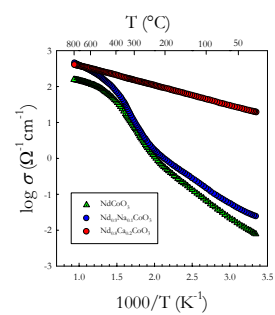

Lattice effects and transport properties of orthorhombic distorted perovskite-type $NdCoO_3$ based materials have been investigated. The high conductivity at low temperatures, together with the linear thermal evolution of unit cell volume, makes the $Nd_{0.8}Ca_{0.2}CoO_3$ compound a more attractive material than pure $NdCoO_3$ or Na-doped compound.

# 1. INTRODUCTION

Rare earth cobaltates of general formula $Ln_{1-x}A_xCoO_3$ (Ln = rare earth, A = alkaline or alkaline earth metal) show interesting structural, magnetic and transport properties, which are sensitive to the average ionic radius of the A-site cations and are closely related to the availability of different oxidation and spin states for the Co ion. The research interest in cobalt containing oxides ranges over a wide variety of fields, from sensor devices and catalysts that oxidize CO and $CH_4$ or reduce NO [1-10] to components in solid oxide fuel cells [11], from the unusual thermoelectric behaviour [12-14] to the peculiar magnetic properties [15-16], up to the latest discovery of superconductivity in layered systems such as $Na_{0.3}CoO_2 \cdot 1.3H_2O$ [17,18].

It is well known that in compounds possessing the perovskite structure Co ions can exist in three different spin states, low spin (LS), intermediate spin (IS) and high spin (HS) because the crystal field splitting of the Co-$d$ states and Hund's rule coupling energy are comparable for these compounds. This implies that thermal energy can induce the spin-state transition of Co ions by thermal excitation of the $t_{2g}$ electrons into $e_g$ states and this spin-state transition has represented a rather important aspect largely studied on $LaCoO_3$ [19,20]. More recently, an analogous spin-state transition has been characterised also in cobaltates of Pr and Nd [20,21], showing that the transition temperature varies as a function of the rare earth ion in the $LnCoO_3$ series.

Both external applied pressure and internal chemical pressure can induce a spin state transition in Co containing perovskite. A pressure induced IS to LS transition has been observed in $LaCoO_3$ at approximately 4 GPa [23]. Consistently, the replacement of $La^{3+}$ by a smaller lanthanide ion can result in an internal chemical pressure similar to that of a physical pressure on the Co-O bonds, and stabilize the LS configuration [21]. Previous works [21,22] have established that $Co^{3+}$ is predominantly in its LS state in $NdCoO_3$ at room temperature and ambient pressure. Structural phenomena can be associated to the spin-state transition and have been observed in $LaCoO_3$ [24] since, for example, the LS $Co^{3+}$ is an inactive Jahn-Teller ion while the IS $Co^{3+}$ is strongly Jahn-Teller active. An important aspect closely linked to the transport properties of Co containing oxides is the possibility of charge disproportionation of the Co ions; studies of the electrical properties of several oxides have been interpreted based on this phenomenon [25]. In general, among transition metal oxides, charge disproportionation is most prevalent for transition metal ions with partially filled 3$d$ subshell. Due to the possibility of different spin states for $Co^{3+}$, charge disproportionation is, therefore, expected for IS or HS $Co^{3+}$ ions but not for the LS configuration.

The physical properties of perovskites and their possible use in applications strongly depend upon doping effects on the A site that change the average ionic radius on this site and introduce point defects for charge compensations. It is well known that alkaline-earth doped $LaCoO_3$ and $NdCoO_3$ are widely used materials for applications as catalysts, sensor and thermoelectric materials [4-6, 14]. The introduction of a small amount of dopant ions, usually around 20%, introduces a correspondent amount of extrinsic oxygen vacancies which represent, for example, the available adsorption sites for catalysis or increase the ionic contribution to the electrical conductivity.

In this study, the effect of cation doping on the structural and electrical properties of $NdCoO_3$ is analysed as a function of temperature by means of high temperature x-ray and neutron diffraction, thermoelectric power and direct-current (DC) electrical conductivity measurements. One of the most notable and relevant aspects of this work is that, unlike previous structural works [24], the diffraction measurements have been collected under the same experimental conditions as the transport properties measurements. This implies that the correlation between structural and electrical properties that we can obtain from these sets of data are highly reliable and meaningful. Finally, the overall scenario we derived from this measurement is interpreted in view of the different spin-states of the samples considered which has been confirmed by magnetic measurements. We stress that this study is performed on doped samples, the ones effectively important in view of possible applications. Results obtained for doped samples are compared to the pure $NdCoO_3$ material so as to obtain valuable information on the possible improvements introduced by doping.

## 2. EXPERIMENTAL

Powder samples of $NdCoO_3$, $Nd_{0.9}Na_{0.1}CoO_3$ and $Nd_{0.8}Ca_{0.2}CoO_3$ were prepared by conventional solid state reaction from stoichiometric amounts of $Nd_2O_3$, $Co_3O_4$, $CaCO_3$ and $Na_2CO_3$ (all from Aldrich Chemicals, ≥99.9% purity) by repeated grinding and firing for 24 hours at temperatures between 900 and 1050 °C. The amount of the two dopant ions has been chosen in order to ensure the same oxygen vacancies concentration for the two compositions, in agreement with the different oxidation state of the aliovalent dopants. Room temperature X-ray powder diffraction (XRPD) patterns were collected on all samples to ensure the phase purity using a Bruker-D8 Advance Diffractometer employing Cu anticathode radiation.

Total conductivities of the samples were measured, using four probe DC-conductivity techniques, from room temperature up to 750°C. Rectangular-shaped pellets of each sample were sintered at 950 °C for 48 h and coated at both short ends with silver paste to provide electrical contacts for the measurements. Temperature dependent thermoelectric power measurements were also carried out under the same conditions as the conductivity measurements. They consisted in measuring the electromotive force created by a temperature gradient across the sample in open circuit and recorded by means of two distinct thermocouples.

The Co-K edges XAS spectra were collected in transmission mode at room temperature at the BM29 beamline of the ESRF synchrotron radiation laboratory (Grenoble, France) using ion chambers as detectors. For all the measurements, the samples were mixed with cellulose and pressed into pellets. The amount of sample in the pellets was adjusted to ensure a total absorption ($\mu$) above the edge around 2. Spectra were processed by subtracting the smooth pre-edge background fitted with a straight line. Each spectrum was, then, normalized to unit absorption at 1000 eV above the edge, where the EXAFS oscillations were no longer visible.

The high temperature (30°C – 750°C) X-ray powder diffraction (XRPD) measurements have been carried out at the Swiss Light Source at the Swiss Light Source Materials Science beamline powder diffraction station. The samples were loaded in quartz capillaries and the measurements carried out at ambient pressure in the temperature range between room temperature and 700°C, approximately every 20°C. 60 degrees full diffraction patterns were collected using the fast MYTHEN microstrip detector at a wavelength of λ=0.70855 Å, as determined using a standard silicon powder (NIST 640c) [34].

Neutron powder diffraction data were acquired on the D2B instrument at the Institute Laue Langevin (ILL) in Grenoble. All measurements were recorded in air in a silica glass container and the diffraction patterns collected in the angular range 0°-160°, step 0.05°, wavelength 1.59 Å, for 6 hours on the D2B instrument in the temperature range between room temperature and 650°C every 100°C. Analogously to the XRPD acquisition, the ND measurements have been carried out in an "open" furnace, thus avoiding the use of vacuum which would have led to the samples reduction.

All the neutron and the X-ray diffraction patterns were analysed according to the Rietveld method [26,27] by means of the FULLPROF software package [28]. For the XRPD patterns the background was fitted with an interpolation between fixed point chosen outside Bragg peaks. Cell parameters, atomic position and isotropic thermal factors for all the ions were refined. For the NPD patterns the background coming from the empty quartz tube (recorded at the same temperatures of the samples) was subtracted from the NPD patterns of the samples. Typical $R_{wp}$ values for the NPD (XRPD) refinements are around 7-12 (8-12) with $\chi^2$ values between 1.2 and 2 (1.5-2) for all the temperatures and samples.

Thermogravimetry (TGA) measurements carried out under the same conditions of the diffraction measurements showed weight changes of the order of ~0.02% for all the samples in the $T$ range explored. This allows us to safely conclude that the weight change observed does not influence the Co valence state as a function of temperature.

Complementary magnetization measurements have been performed at room temperature with a QD Squid magnetometer, with magnetic field ranging between 0 and 7 T.

# 3. RESULTS AND DISCUSSION

*3.1 Transport Properties*

The thermal evolution of the total conductivity of pure (NCO), Na-doped (NNCO) and Ca-doped (NCCO) NdCoO$_3$ compounds is presented in Figure 1-A in the form of an Arrhenius plot. The overall conductivity behaviour of the three samples is clearly different. In particular, the total conductivity values at room temperature are: ~20 $\Omega^{-1}$cm$^{-1}$ for NCCO, ~2×10$^{-2}$ $\Omega^{-1}$cm$^{-1}$ for NNCO and ~7×10$^{-3}$ $\Omega^{-1}$cm$^{-1}$ for NCO. The conductivity of the pure and the Na-doped samples is characterized by a non-linear behaviour; a clear slope change in the temperature range around 230–400°C indicates an increased activation energy for the two samples in this temperature range.

Figure 1-B shows the thermoelectric power for the three samples. At room temperatures they all have a high positive value that rapidly decreases with increasing temperature and becomes small in magnitude and nearly temperature independent above ~400°C, where the three curves converge asymptotically at the same positive value indicating a *p*-type conductivity. The reduction of the thermoelectric power by increasing the temperature is connected to the increased number of charge carriers according to the Heikes's formula [33]:

$$S = \frac{k}{e} \ln\left|\frac{1-c}{c}\right| \qquad (1)$$

where in this case we may consider *c* equal to a small polaron or a Co$^{4+}$ ion. This effect is in turn closely connected to a spin-state transition from LS to IS or HS species that are the only spin states able to create Co$^{4+}$ ions through the disproportionation mechanism [25,29]. The differences in the value of the low-*T* thermoelectric power can be attributed either to differences in the relative population of Co spin-states among the three samples or in the density of state at the Fermi level as a consequence on the extension of the conduction band. In this case it looks that a strong localization of the charge carriers is triggered by the Na-doping. In the high temperature limit all the samples show approximately the same amount of charge carriers suggesting the presence of a common spin-state and $Co^{4+}$ population.

The comparison of the samples presented throughout the paper gains a high significance not only because all the experimental evidences are collected under the same conditions but also because the three samples possess the same average Co valence state. This has been determined through Co-K X-ray adsorption spectroscopy (XAS) measurements at room temperature. Figure1-C shows the derivative of the absorption coefficient *vs*. temperature for the three samples; as can be seen, the maximum of the three curves is placed at the same position. Although the lack of a reference standard for the Co oxidation state within a proper matrix (the matrix effect of the binary oxides have proved to make them unusable standards) did not allow us to directly establish the effective value of the average oxidation state, by means of site occupancy refinements of the diffraction patterns we could indirectly extract this information (see Section 3.2).

## *3.2 Structural Properties*

Both the XRDP and NPD diffraction patterns of all samples at all temperatures can be unequivocally indexed in the orthorhombic space group *Pnma*. Figure 2 shows at this regard the refined XRPD and NPD experimental patterns for $NdCoO_3$ (empty circles) together with the calculated profiles (red line), the residuals (blue line) and the theoretical peaks for the orthorhombic perovskite phase (vertical grey lines). In Table 1 the room temperature structural data derived from Rietveld refinement of the XRPD for the three samples are listed. Rietveld refinement of the neutron diffraction patterns have taken into account possible partial occupancies on the oxygen sites. Results on the doped compounds reveal that the refined oxygen stoichiometries are in good agreement with the theoretical values calculated according to the following quasi-chemical equilibria in Kroger-Vink notation:

$$CaO + Nd_{Nd}^{\times} + \tfrac{1}{2}O_O^{\times} \rightarrow Ca_{Nd}' + \tfrac{1}{2}V_O^{\cdot\cdot} + \tfrac{1}{2}Nd_2O_3 \qquad (2)$$

$$\tfrac{1}{2}Na_2O + Nd_{Nd}^{x} + O_{O}^{x} \rightarrow Na_{Nd}^{''} + V_{O}^{\cdot\cdot} + \tfrac{1}{2}Nd_2O_3 \tag{3}$$

Based on the NPD results, the oxygen content for the NCCO sample is 2.87(2) while for the NNCO it is 2.85(3). This allows us to calculate the corresponding Co valence states to 2.94(4) and 2.90(6) respectively. For the NCO sample the refinement converged towards a full occupancy of the two O-sites. This result is particularly significant since it implies that the cation doping does not produce any $Co^{4+}$ ion as often hypothesized in literature. The valence state of the Co ions tends instead to keep a value close to +3 or slightly lower. This behaviour is substantially different that of the manganite perovskites where the stable Mn valence state tends to be a mixed +3/+4 oxidation and the defect chemistry involves the compensation of the aliovalent doping with electronic defects, thus keeping the average oxygen content higher than 3.0. [30,31]. Finally, these results confirm the similarities of the Co valence state between the three samples determined through the XAS and indicate that this valence state is very close to +3.

The thermal evolution of the pseudo-cubic unit cell parameters ($a_p=a/\sqrt{2}$; $b_p=b/2$; $c_p=c/\sqrt{2}$), as derived from synchrotron x-ray diffraction, is presented in Figure 3. In the inset the pseudo-cubic unit cell parameters derived from the refinement of the neutron powder patterns are reported for comparison and provide the evidence for a remarkable agreement between the two sets of results.

By looking at these trends, it is first of all interesting to note the similarity between the pure and the Na-doped compound. For these two samples, at low temperature, *a* is smaller than *c*, with *a* and *c* being the two short axis in this crystallographic setting. For $NdCoO_3$ and $Nd_{0.9}Na_{0.1}CoO_3$ the thermal expansion along the three axis is clearly anisotropic: the *a* axis expands more than the others, so that around 400°C a crossing between *a* and *c* takes place; at lower temperature (approximately 280°C) *a* also crosses the pseudo-cubic cell parameter *b* (the long crystallographic axis in this crystallographic setting).

The behaviour with *T* for the $Nd_{0.8}Ca_{0.2}CoO_3$ compound is different. The pseudo-cubic *c* parameter is here the longest one in the whole temperature range considered; no crossing with the two short lattice parameters is visible, while the crossing between the pseudo-cubic *a* and *b* lattice parameters takes place at a higher temperature (about 400°C). Figure 4-A shows the trend of the cell volume for the three samples as a function of temperature. The values of the unit cell volumes of the three sample are, at room temperature, very close to each other, in agreement with the small differences in the average ionic radius on the A site for the three samples, *i.e.*, $NdCoO_3$ ($<r_A>$ = 1.270 Å), $Nd_{0.9}Na_{0.1}CoO_3$ ($<r_A>$ = 1.282 Å), $Nd_{0.8}Ca_{0.2}CoO_3$ ($<r_A>$ = 1.284 Å). The unit cell volume of the pure and Na-doped compounds is almost equal. Although, based on the difference in the average ionic radius on the A site, we would expect a higher value for the Na-doped sample, we have to qualitatively take into account the effect of oxygen vacancies.

The trends shown in Figure 4-A reveal three main points:

- $Nd_{0.8}Ca_{0.2}CoO_3$ expands almost linearly with temperature, while $NdCoO_3$ and $Nd_{0.9}Na_{0.1}CoO_3$ exhibit a non-linear expansion behaviour with temperature;
- between ~250 and ~500°C the unit cell volume of $NdCoO_3$ and $Nd_{0.9}Na_{0.1}CoO_3$ expands more rapidly than in the other temperature ranges;
- due to its linear expansion with temperature, the unit cell volume of $Nd_{0.8}Ca_{0.2}CoO_3$ is considerably smaller than that of the other two samples at high temperature.

The overall unit cell distortion of the three samples has been quantified through a distortion parameter (*d.p.*) calculated as following:

$$d.p. = \frac{\sqrt{\sum (a_i - \bar{a})^2}}{\bar{a}} \qquad (4)$$

where $a_i = a/\sqrt{2}, b/2, c/\sqrt{2}$ and $\bar{a}$ is the average of $a_i$. According to this definition, the unit cell distortion reaches a minimum at *d.p.* = 0 when the three pseudo-cubic lattice parameters are equivalent. Figure 4-B shows the evolution with temperature of this parameter for the three samples. The main considerations coming from these results are the following:

- at room temperature $Nd_{0.8}Ca_{0.2}CoO_3$ is considerably less distorted than the other two samples.

- the distortion parameters reaches a minimum for $NdCoO_3$ and $Nd_{0.9}Na_{0.1}CoO_3$ at ~380 and ~400°C as confirmed by the analysis of the thermal evolution of the unit cell parameters (see Figure 3).

- the distortion of the $Nd_{0.8}Ca_{0.2}CoO_3$ unit cell is approximately constant and lower than that of the other samples up to approximately 250°C; then it starts increasing continuously with $T$. From 250°C on, the distortion of the Ca-doped compound is higher than for the other two samples.

Although the unit cell parameters of $NdCoO_3$ and $Nd_{0.9}Na_{0.1}CoO_3$ eventually become very close to each other, there is no evidence of a phase transition to a more symmetric system.

The degree of distortion of a perovskite compound is usually quantified by means of the well-known Goldschmidt tolerance factor ($t$). The geometrical tolerance factor is defined as follows:

$$t = \frac{\langle A-O \rangle}{\sqrt{2}\langle B-O \rangle} \tag{5}$$

where <A-O> represents the average cation-oxygen distance on the A site and <B-O> the average cation-oxygen distance on the B site of the perovskite structure. The equilibrium bond lengths are usually calculated for ambient conditions from the sums of the ionic radii in the appropriate coordination numbers for the atoms considered. The tolerance factor is a function of temperature and pressure: a larger thermal expansion of the (A–O) bonds makes $dt/dT > 0$ except where a spin-state transition occurs on the B cation. [32] As $t$ approaches 1, the structure is expected to become less distorted, $t=1$ being the situation of the ideally cubic perovskite; the distortion associated with a $t$ smaller than 1 is accommodated by the cooperative rotation of $BO_6$ octhaedra, which lower the symmetry of the structure. Doping on the Nd site with a larger cation increases the effective mean ionic radius of this site, thus increasing $t$ at ambient conditions. It is worth remembering here that a reduction in the distortion of the structure implies a reduction of both the overall unit cell distortion and the "local" distortion, mainly due to the internal distortion of the $CoO_6$ octahedra (bond lengths) and their bending angles Co-O-Co.

In orthorhombic perovskites like $NdCoO_3$, there are two Co-O-Co angles, in agreement with the presence of two distinct O sites (indicated as O1 and O2), as shown in Figure 5. While the Co-O2-Co bending is responsible for the rotation of the $CoO_6$ octahedra with respect to the *ac* plane, the Co-O1-Co angle is a measure for a deviation from the ideal value of 180° along the *b* axis. Within the $CoO_6$ octahedra, three different Co-O distances can be distinguished; in particular, as shown in Figure 5, one Co-O1 bond length in apical position and two Co-O2 in the equatorial plane. Depending on the relative differences between these three bonds, the Co environment is more or less distorted.

In Figure 6, the $Co-O_i$ distances *vs.* temperature trends are reported for the pure (NCO) and the Ca-doped (NCCO) compounds chosen as meaningful comparison between two samples with completely different structural and electrical behaviour (as already seen for the unit cell, the Na-doped sample behaves in an analogous way as the pure compound). In the upper panel (Figure 6-A and Figure 6-B) is reported the comparison between the Co-O1 and the average Co-O2 bond lengths for the two samples as a function of temperature while in the lower panel (Figure 6-C and Figure 6-D) the trend with *T* for the three $Co-O_i$ distances is shown.

The relative difference between apical and equatorial $Co-O_i$ bond lengths gives an indication of the possible octahedral Jahn-Teller distortion and its evolution with temperature. It is, therefore, interesting to note that while for $Nd_{0.8}Ca_{0.2}CoO_3$ the single apical Co-O1 and the average equatorial Co-O2 bond lengths are almost equal for all the temperatures considered, due to a linear expansion with temperature, the same parameters have a more complex behaviour for pure $NdCoO_3$. We can split the temperature range considered in three parts: between room temperature and ~250°C the average equatorial Co-O2 bond length is only slightly longer than the apical Co-O1; between ~250°C and approximately 500°C, due to different thermal expansion coefficients of the bond lengths, the apical Co-O1 becomes longer than <Co-O2>, starting from about 380°C the Co-O1 reaches a "plateau" while the <Co-O2> continues to expand linearly so that after 500°C the equatorial bond again becomes longer than the apical one. It should be noted that while there is no clear out-of-plane Jahn-Teller distortion above 500°C, the separation between the two equatorial Co-O2 bond lengths above this temperature is indicative of an octahedral distortion in the *ac* plane. At higher temperature the two Co-O2 bond lengths start again converging, thus reducing the equatorial distortion. A different behaviour is observed for $Nd_{0.8}Ca_{0.2}CoO_3$, where the three Co-Oi distances are different at low temperature and converge at the same value above ~500°C.

Figure 7 shows the behaviour of the Co-Oi-Co angles *vs.* temperature. Again, the behaviour is considerably different for the two samples. The average Co-Oi-Co angle is approximately independent on temperature for $Nd_{0.8}Ca_{0.2}CoO_3$ with a mean value of approximately 158° and there is at the same time a substantial equality between the Co-O1-Co and Co-O2-Co angles. The situation is more complex for the pure compound (Figure 7-A). The average Co-Oi-Co angle is basically constant at a value of ~157.5° from room temperature up to ~250°C; then it rapidly decreases, reaching a minimum of ~155° at ~380°C and, finally, it returns to the initial value. As shown in Figure 7-C this behaviour is the result of two opposite effects: an abrupt drop of the Co-O1-Co angle between 250 and 500°C, with a minimum of about 153° at 380°C, and a simultaneous increase in the same temperature range of the Co-O2-Co angle, reaching a maximum of ~159° at 300°C. The overall result again suggests that it is possible to identify for the NCO compound (and similarly for the Na-doped sample) three temperature ranges: between room temperature and ~250°C where the Co-O1-Co angle is more open than the Co-O2-Co tilting angle, between ~250 and ~500°C where the situation reverses (with a significant difference of ~6° at ~380°C) and above ~500°C where the first situation is restored.

Finally, Figure 8 shows the evolution with temperature of the tolerance factor, calculated by means of the refined bond lengths distances, according to equation (5). Again, we find a significantly different behaviour between $NdCoO_3$ and $Nd_{0.8}Ca_{0.2}CoO_3$; in particular, the negative slope of the tolerance factor *vs*. temperature for the NCO in the approximate temperature range 250-400°C and the similarity of these trends to the changes with temperature of the Co-O-Co angles.

From the combined analysis of all these results, it is clear that $NdCoO_3$ and $Nd_{0.9}Na_{0.1}CoO_3$ have an analogous thermal behaviour, both in the structural and in the electrical characterization, in contrast to the $Nd_{0.8}Ca_{0.2}CoO_3$ sample. The similarity between the first two samples and the excellent agreement between the structural results coming from the neutron and x-ray diffraction measurements suggests to only consider $NdCoO_3$ and $Nd_{0.8}Ca_{0.2}CoO_3$ in the following discussion.

Three temperatures are of particular relevance throughout this study: ~250, ~380 and ~500°C suggesting to divide the overall temperature range in three regions: between 30 and 250°C, between 250 and 500°C and between 500 and 750°C.

First of all we observe that a strong correlation exists between the various lattice effects and the transport properties as a function of temperature for $NdCoO_3$ and $Nd_{0.9}Na_{0.1}CoO_3$: when a change in the cell volume expansion occurs, a change in the bond-lengths and angles *vs*. temperature also occurs, and a slope change in the Arrhenius plot is visible. Furthermore, it is possible to correlate these changes with a clear slope change in the tolerance factor *vs*. temperature plot. We now clarify these correlations that, again, we believe are highly valuable due to the fact that, to the best of our knowledge, this is the first time that lattice effects and transport phenomena on a cobalt containing perovskite are studied under exactly the same experimental conditions. In particular we demonstrate that it is possible to correlate the structural changes and the transport behaviour on the basis of a spin state transition. We present evidences supporting the sequence LS–IS–HS for $NdCoO_3$ and the sequence IS–HS for $Nd_{0.8}Ca_{0.2}CoO_3$, taking into account also the possibility of a high temperature induced disproportionation reaction for the Co ions. In fact different spin states of the Co ions correspond to different average ionic radius on the B site ($r_{Co3+}^{LS}$ = 0.545 Å; $r_{Co3+}^{IS}$ = 0.56 Å; $r_{Co3+}^{HS}$ = 0.61 Å) and to different Jahn-Teller properties, so that the Co environment is a really sensitive probe of the local distortion view in terms of bond-lengths and bending angles.

A spin state transition between LS $Co^{3+}$ and IS $Co^{3+}$ can account for all the "local" structural distortions and their changes as observed for the pure compound. Below 250°C there is no clear Jahn-Teller effect, as the three Co-Oi bond lengths have practically the same value (Figure 6-C). This is consistent with the presence in this temperature range of a large majority of LS $Co^{3+}$ ions, a Jahn-Teller inactive ion, in agreement with previous experimental work [20,22]. The transition to a higher Co spin state, in particular to the intermediate Jahn-Teller active spin state IS, is then reflected in the occurrence of a static Jahn-Teller distortion at higher temperature (Figure 6). The difference in ionic radii between the LS and IS $Co^{3+}$ ion, could justify the increased thermal expansion coefficient above 250°C observed in Figure 4-A. Indeed, if the smaller LS $Co^{3+}$ ions is progressively substituted by the larger IS $Co^{3+}$ ion starting from about 250°C we would expect, in addition to the normal thermal expansion that would results in a linear unit cell volume *vs.* temperature trend, an additional term due to the introduction of the larger IS cation on the B site The increase in the average ionic radius of the B site is also consistent with the reduction of the tolerance factor (Figure 8) and the corresponding increase in the distortion of the structure. Above 250°C the $CoO_6$ octhaedra are more distorted since by looking at Figure 6-A it is a clear that the internal Co-Oi bond lengths are different, while the tilting of same octahedra along the *b* axis sensibly increases since the Co-O1-Co angle decreases by about 5° (Figure 7-C). Interestingly, the cooperative effect of this bond length and angle distortion produces the favourable conditions for the unit cell parameters to become closer to each others, so that at ~380°C we observe a crossing between the two short *a* and *c* lattice parameters. Consistently, as the Co-Oi-Co angles decrease, the activation energy in the Arrhenius plot (Figure 1-A) increases, as it is well known that in a small polaron system the hopping mechanism at the basis of the electrical conductivity is favoured when the hopping angle (in this case Co-O-Co) is closer to 180°. In the same temperature range, the absolute value of the conductivity increases and the thermoelectric power decreases as the number of charge carriers increases as soon as the LS-IS transition occurs. At high temperature, when the Co-Oi distances seem to again converge and the Co-O-Co angle restores its original value, the structure becomes locally less distorted, which is most probably connected to a new spin state transition towards a HS state.

For $Nd_{0.8}Ca_{0.2}CoO_3$, the structural and thermoelectric power data agree with the presence at room temperature of a consistent fraction of IS $Co^{3+}$, supported by the bond-length distortion within the $CoO_6$ octahedra at low temperature (Figure 6-D) and the high conductivity compared to the pure compound. At higher temperature this distortion disappears since the three bond lengths converge at the same value, with again a possible spin state transition towards a HS.

Finally, for the $Nd_{0.1}Na_{0.1}CoO_3$ sample the overall behaviour both in the transport and structural properties closely resembles the one of the pure sample. The higher thermoelectric power value at low temperature and up to the LS-IS spin transition, with respect to that of NCO, most probably witness both a higher fraction of LS Co ions and a higher localization of the charge carrier induced by the Na-doping with respect to the pure sample and in a totally opposite fashion with respect to the Ca-doped sample.

In order to confirm the spin-state characteristics of the samples, which we have derived from the structural data, complementary magnetization measurements have been carried out as a function of the applied field at 300 K. These are shown in Figure 9. The *M vs. H* dependence shows a pure paramagnetic behaviour for all the samples with NCO and NNCO showing a very similar value of the magnetization and NCCO showing a higher value. By comparing the experimental values of the paramagnetic moment for NCO and NNCO samples we can safely attribute the whole magnetic contribution in these two samples to the $Nd^{3+}$ ions. This result is in total agreement with a LS state for the two compositions since the Co ions in the LS configuration do not give any contribution to the total magnetization.

In the case of the NCCO sample the higher contribution to the magnetic susceptibility is reliably attributable to the $Co^{3+}$ ions in an intermediate spin state. As a matter of fact, the "extra" contribution to the *M* value due to the Co ions, in addition to the one of the $Nd^{3+}$ ions, leads to an experimental value of 2.86 Bohr magnetons which is in perfect agreement with the calculated value of 2.83 Bohr magnetons in the case of a ion with $S=1$.

# 4. CONCLUSIONS

In this study we aimed at clarifying the effect of aliovalent cation doping on the structural and transport properties of NdCoO$_3$ as a function of temperature by covering a temperature range of applicative interest for catalyst and sensor activity, i.e. the high-temperature region. The extensive set of data, collected under controlled, reliable and analogous experimental conditions, point out a strong correlation between the lattice effects, derived by high temperature x-ray and neutron diffraction, and the transport properties of the materials investigated by means of conductivity and thermoelectric power measurements. We like to recall here that the three samples investigated have the same Co valence state, thus allowing us to correlate the different behaviours observed to the cation doping effect alone. To the best of our knowledge this is another unique feature of this work with respect to the available literature results.

This work revealed the presence of three relevant temperatures: ~250, ~380 and ~500°C that subdivide the overall considered temperature range in three distinct regions whose nature and characteristics have been deeply highlighted in the previous section.

To summarize, a first important conclusion that can be drawn from this study is the dramatic improvement of the room temperature conductivity of Nd$_{0.8}$Ca$_{0.2}$CoO$_3$ with respect to the pure and the Na-doped compounds. This can not be attributed to the presence of extra-oxygen vacancies, since at this temperature the conductivity is totally electronic, but has to be connected rather to the predicted spin-state for the Co ions within this structure. It has long been believed that the transport properties of perovskite oxide, together with other physical properties of these oxides, could be rationalized in terms of their tolerance factor and, in turn, their degree of distortion, in particular octahedral distortion and tilting. The experimental data presented in this study show that the Ca-doped compound is far less distorted than the pure compound if we consider the overall cell distortion: the measured tolerance factor at room temperature for the Ca-doped compound is slightly higher than NdCoO$_3$. However, this slight difference can not explain on its own the dramatic improvement in the transport properties especially since the local distortion of the Ca-doped sample is even higher than that of the pure compound: the average Co-O distances are approximately the same for the two samples, also the average Co-O-Co angle at room temperature is approximately the same, while the octahedral bond lengths distortion is more pronounced for Nd$_{0.8}$Ca$_{0.2}$CoO$_3$. The main difference between the two samples is more likely related to the Co spin state. The experimental data presented in this study support the presence of a large majority of Jahn-Teller active IS Co$^{3+}$ at room temperature for the Nd$_{0.8}$Ca$_{0.2}$CoO$_3$ and of Jahn-Teller inactive LS Co$^{3+}$ at room temperature for Nd$_{0.9}$Na$_{0.1}$CoO$_3$ and NdCoO$_3$. The presence of IS Co$^{3+}$ ions at room temperature for Nd$_{0.8}$Ca$_{0.2}$CoO$_3$ accounts for the high conductivity of this sample and for the lower Seebeck coefficient in comparison to the pure compound. In fact, IS Co$^{3+}$ can directly participate in the hopping mechanism and has the greatest tendency among the different Co$^{3+}$ spin states to form Co$^{4+}$ species as a consequence of a disproportionation reaction. The available structural data support the presence of a thermally activated LS–IS transition for Nd$_{0.9}$Na$_{0.1}$CoO$_3$ and NdCoO$_3$. The Jahn-Teller inactive LS Co$^{3+}$ present at room temperature progressively transforms in the Jahn-Teller active IS Co$^{3+}$ with increasing temperature starting form approximately 250°C. While this change occurs, the unit cell volume expands more rapidly, in agreement with the difference in ionic radius between the two species, and for the same reason the tolerance factor decreases. The conductivity values increase, consistently with the promotion of new charge carriers, although also the activation energy increases, in correspondence of the reduction of the hopping angle. In the high temperature range the conductivity value, the activation energy and the Seebeck coefficient are independent upon the composition. In order to obtain the same average number of charge carrier and similar conditions for transport, the structure can accommodate a consistent amount of changes. Although based on the available data it is difficult to establish if a possible transition to a HS Co$^{3+}$ state is more likely to occur before or in concomitancy with the Co disproportionation, the present study has underlined the strict correlations between structural and transport properties of these materials, originating by the ability of the Co ions to posses different spin and oxidation states.

In view of possible applications of these materials in any devices, it is highly advisable to take in consideration their thermal evolution and the role of the peculiar dopant atom used. The high conductivity at room temperature, together with the linear thermal evolution of unit cell volume and total conductivity makes the $Nd_{0.8}Ca_{0.2}CoO_3$ compound, for the same Co valence state, a more attractive material than the pure $NdCoO_3$ or Na-doped compound.

# TABLES

**Table 1** Room temperature structural data for $NdCoO_3$ (NCO), $Nd_{0.9}Na_{0.1}CoO_3$ (NNCO) and $Nd_{0.8}Ca_{0.2}CoO_3$ (NCCO) as derived form XRPD refinement

|       |   | NCO         | NNCO        | NCCO        |
|-------|---|-------------|-------------|-------------|
| $a$ (Å) |   | 5.33463 (1) | 5.33379 (1) | 5.34340 (1) |
| $b$ (Å) |   | 7.55217 (1) | 7.55190 (1) | 7.55400 (2) |
| $c$ (Å) |   | 5.34838 (1) | 5.34819 (1) | 5.34910 (1) |
| $V$ (Å$^3$) |   | 215.476 (1) | 215.426 (1) | 215.911 (1) |
| Nd/M  | $x$ | 0.03425 (5) | 0.03422 (5) | 0.03272 (6) |
|       | $y$ | 0.25        | 0.25        | 0.25        |
|       | $z$ | -0.0060 (1) | -0.0055 (1) | -0.0053 (2) |
|       | B | 0.67 (1)    | 0.62 (1)    | 0.76 (1)    |
| Co    | $x$ | 0.00        | 0.00        | 0.00        |
|       | $y$ | 0.00        | 0.00        | 0.00        |
|       | $z$ | 0.00        | 0.00        | 0.00        |
|       | B | 0.75 (1)    | 1.0 (1)     | 0.67 (1)    |
| O1    | $x$ | 0.4909 (7)  | 0.4924 (7)  | 0.4904 (7)  |
|       | $y$ | 0.25        | 0.25        | 0.25        |
|       | $z$ | 0.067 (1)   | 0.070 (1)   | 0.068 (1)   |
|       | B | 0.4 (1)     | 1.2 (1)     | 0.8 (1)     |
| O2    | $x$ | 0.2860 (8)  | 0.2862 (8)  | 0.2811 (9)  |
|       | $y$ | 0.0372 (6)  | 0.0353 (6)  | 0.0364 (6)  |
|       | $z$ | 0.7130 (8)  | 0.7141 (7)  | 0.7146 (8)  |
|       | B | 0.56 (8)    | 1.02 (9)    | 0.56 (8)    |

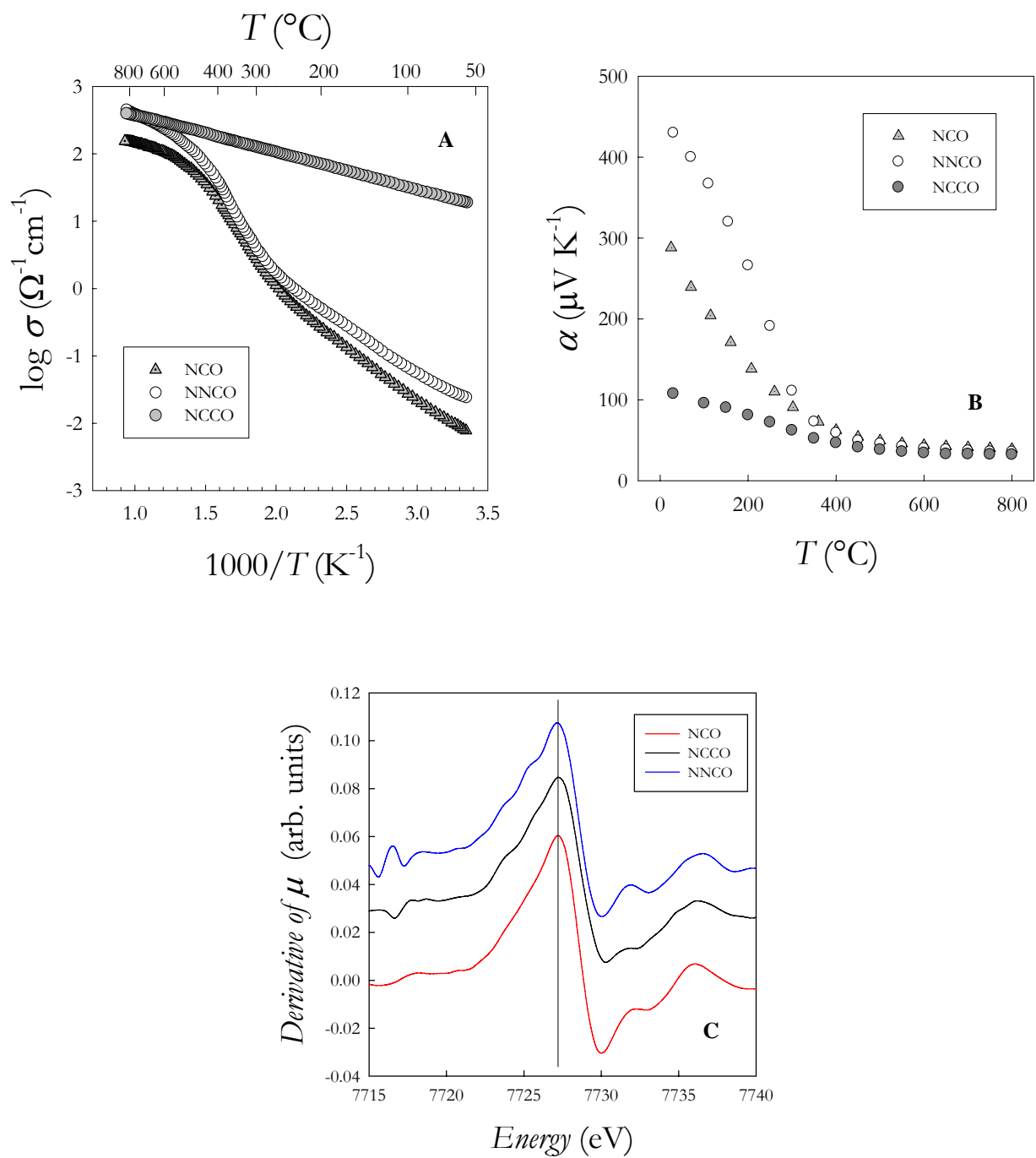

**Figure 1** – Arrhenius plot (A) and thermoelectric power (B) as a function of temperature for $NdCoO_3$ (NCO), $Nd_{0.9}Na_{0.1}CoO_3$ (NNCO) and $Nd_{0.8}Ca_{0.2}CoO_3$ (NCCO). C: Derivative of the absorption coefficient ($\mu$) at the Co-K edge for the three samples in the region around the absorption edge.

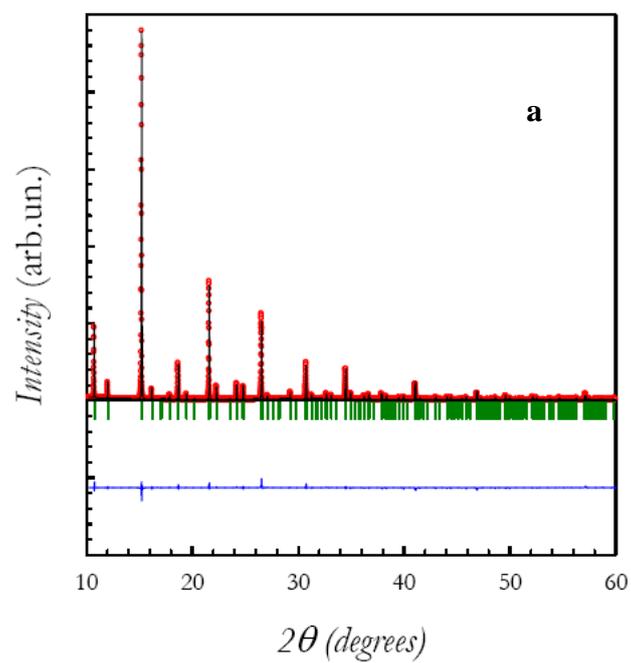

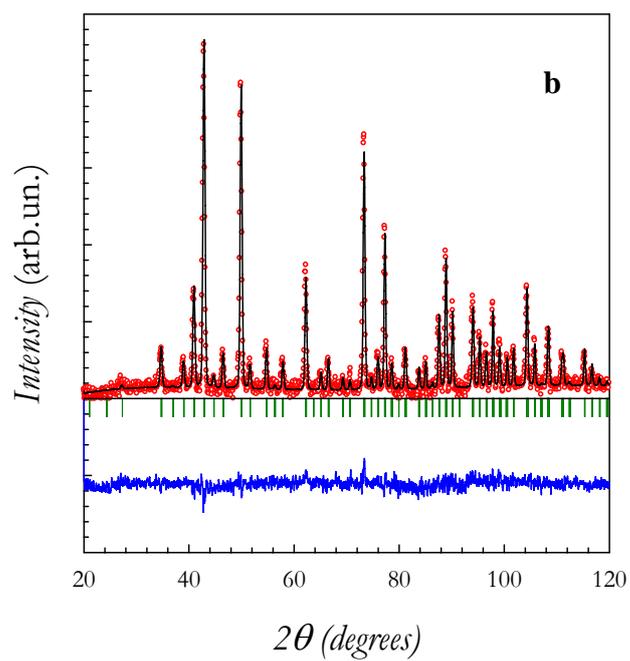

**Figure 2** – Rietveld refined x-ray diffraction (a) and neutron diffraction (b) patterns for NdCoO$_3$ at room temperature: example of observed, calculated and difference profile.

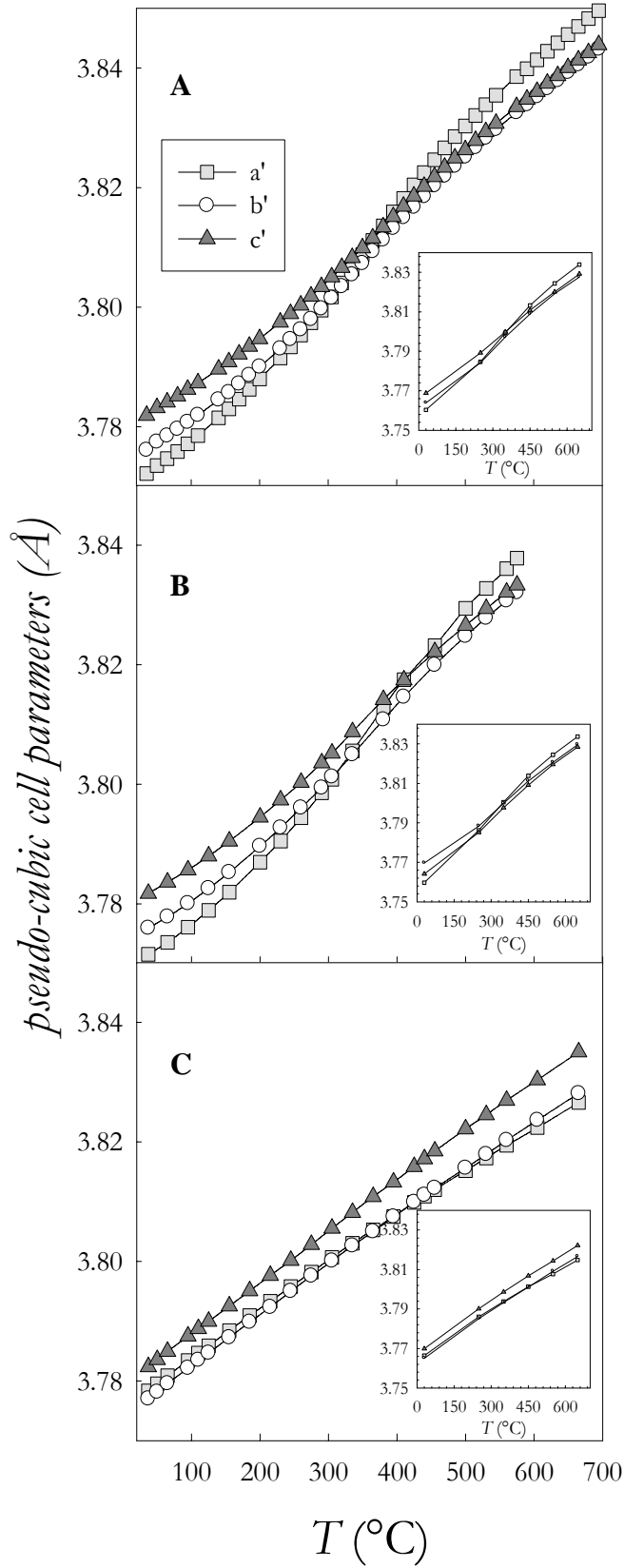

**Figure 3** – Pseudo-cubic cell parameters *vs.* temperature for NdCoO$_3$ (A), Nd$_{0.9}$Na$_{0.1}$CoO$_3$ (B) and Nd$_{0.8}$Ca$_{0.2}$CoO$_3$ (C). Note that the error bars are smaller than the symbol used in the Figure.

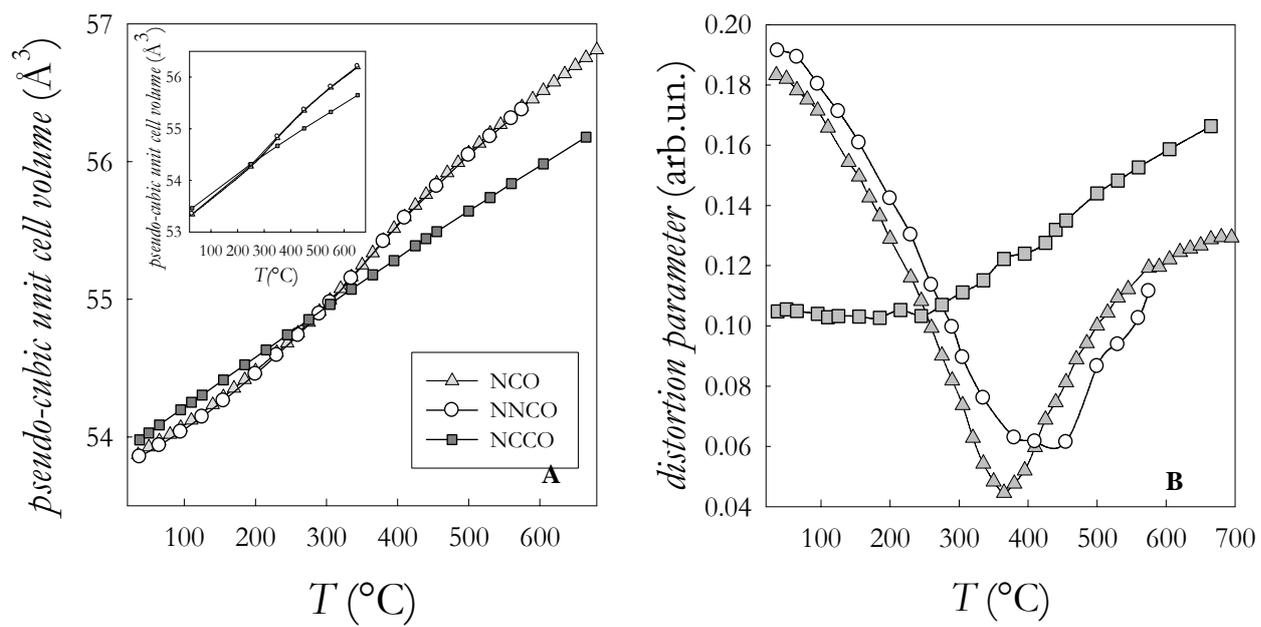

**Figure 4 –** Pseudo-cubic cell volume (A) and cell distortion parameter (B) *vs.* *T* for NCO, NNCO, and NCCO.

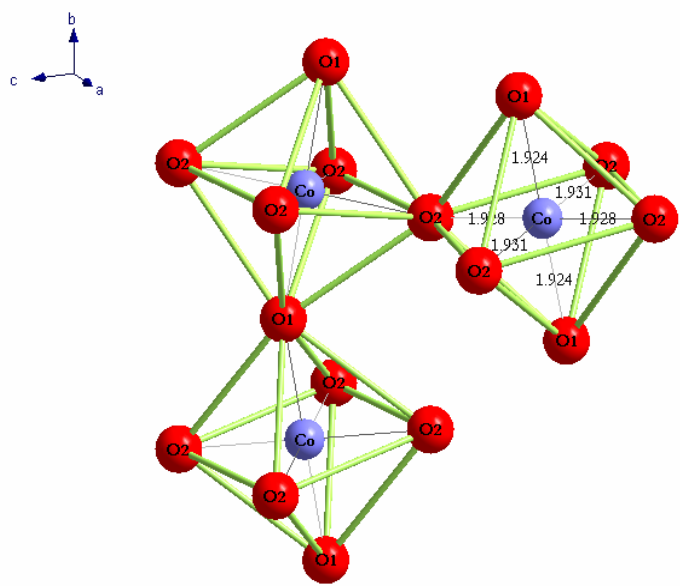

**Figure 5** – Cobalt ion coordination indicating the octahedral distortion and the two Co-Oi-Co tilting angles between adjacent octahedra.

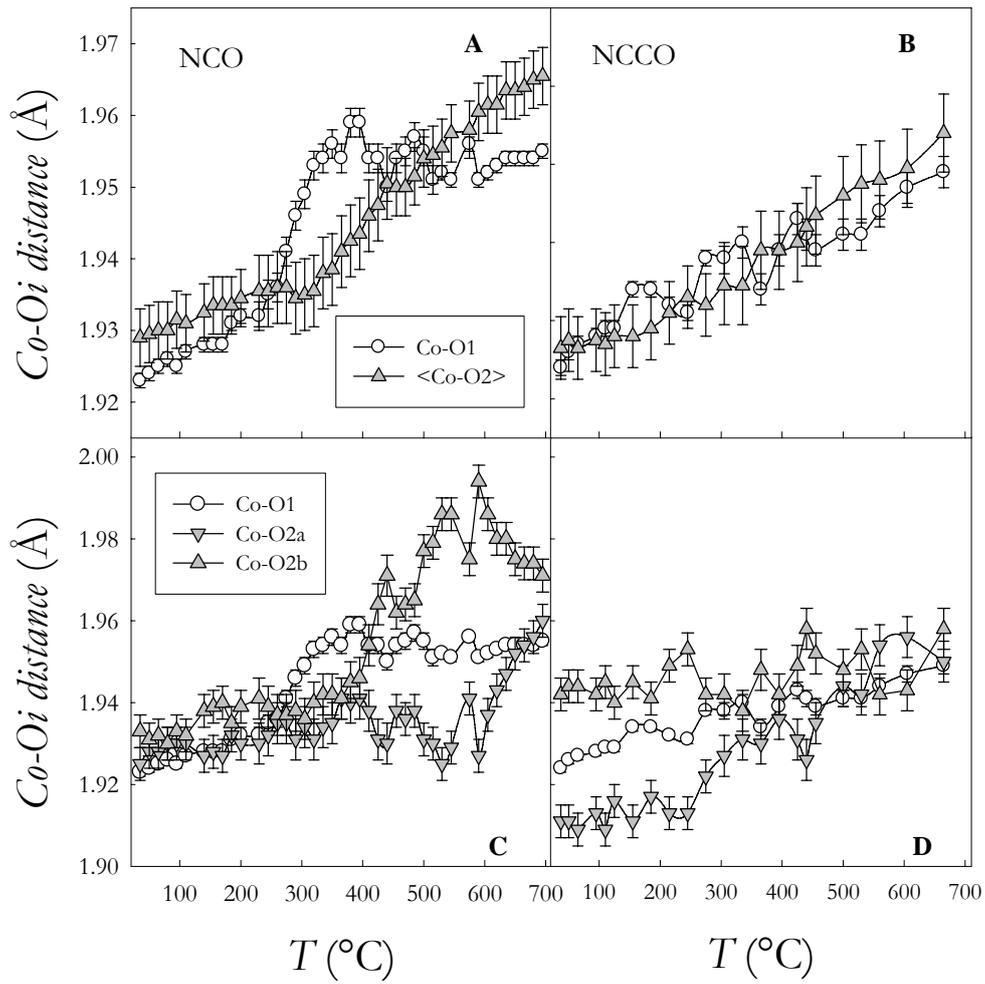

**Figure 6** Co-O bond lenghts evolution with temperature for NCO (A and C) and NCCO (B and D)

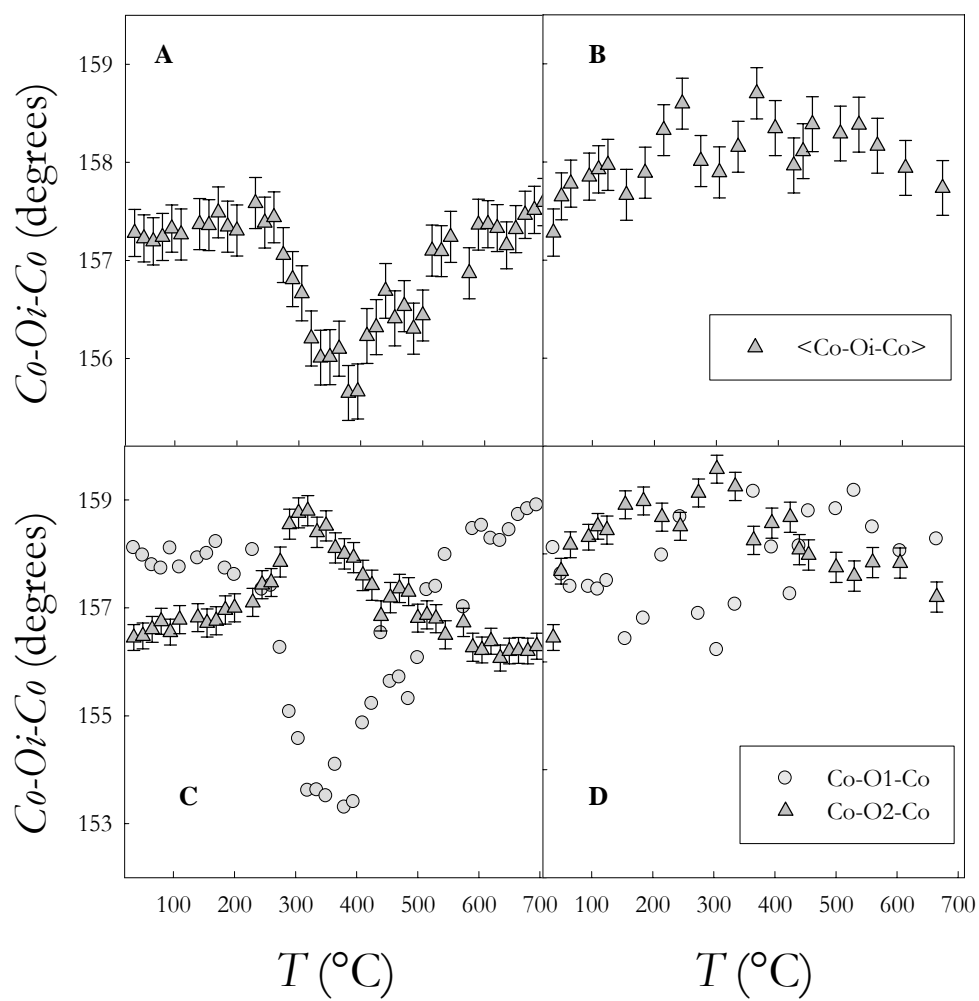

**Figure 7** Bending angles evolution with temperature for NCO (A and C) and NCCO (B and D)

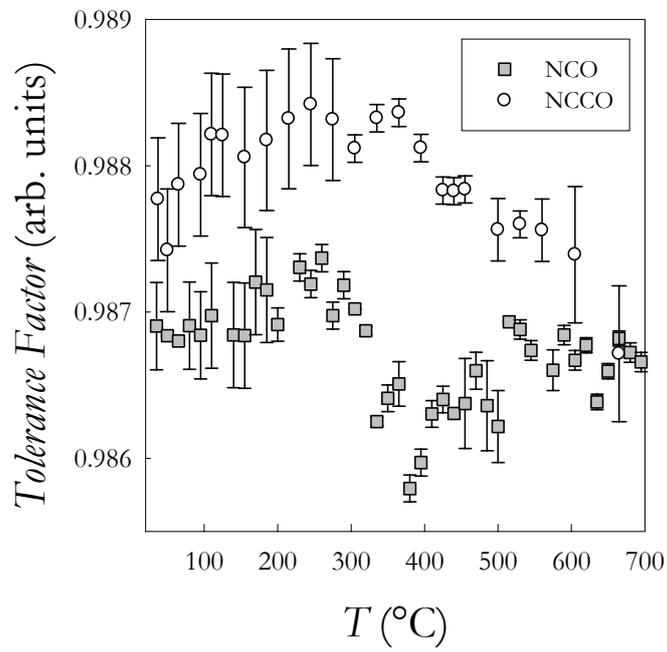

**Figure 8** Evolution with temperature of the geometrical tolerance factor for $NdCoO_3$ (NCO), and $Nd_{0.8}Ca_{0.2}CoO_3$ (NCCO)

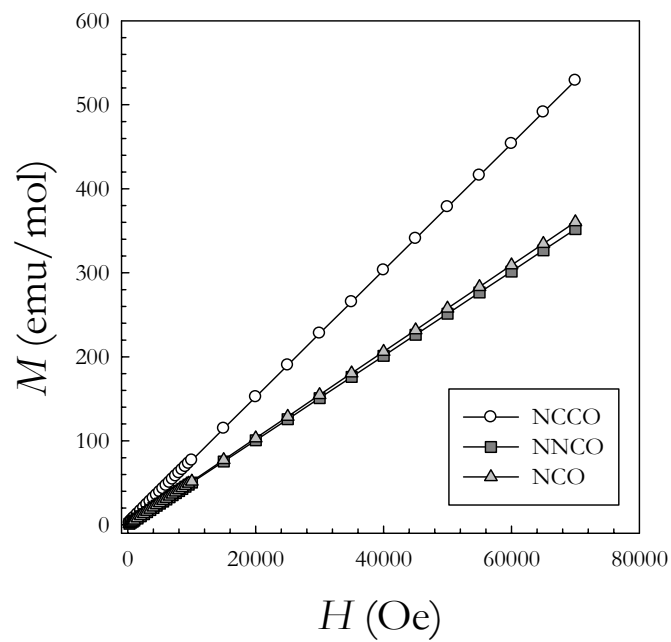

**Figure 9** *M vs H* data at 300 K for NdCoO$_3$ (NCO), Nd$_{0.8}$Ca$_{0.2}$CoO$_3$ (NCCO) and Nd$_{0.9}$Ca$_{0.1}$CoO$_3$ (NNCO).